\documentclass[aps,pra,reprint,twocolumn,superscriptaddress,showpacs,showkeys,longbibliography,groupedaddress]{revtex4-1}

\usepackage{amsmath,amssymb,amstext}
\usepackage[usenames,dvipsnames]{color}
\usepackage{graphicx}
\usepackage{bm,bbold,braket}
\usepackage{natbib}
\usepackage{txfonts, comment,stmaryrd}
\usepackage{dcolumn}

\usepackage[english]{babel}

\usepackage[colorlinks,bookmarks=false,citecolor=blue,linkcolor=red,urlcolor=blue]{hyperref}
\begin{document}

\title{Dipolar condensates with tilted dipoles in a pancake-shaped confinement}

\author{Chinmayee Mishra and Rejish Nath}

\affiliation{Indian Institute of Science Education and Research, Pune 411 008, India}    
\date{\today}

\begin{abstract}
The effect of dipolar orientation with respect to the condensate plane on the mean-field dynamics of dipolar Bose-Einstein condensates in a pancake-shaped confinement is discussed. The stability of a quasi-
two-dimensional condensate, with respect to the tilting angle, is found to be different from a two-dimensional layer of dipoles, indicating the relevance of the transverse extension while characterizing two-dimensional dipolar systems. An anisotropic excitation spectrum exhibiting a highly tunable, rotonlike minimum can arise entirely from the dipole-dipole interactions, by tilting the dipoles. At the magic angle and in the absence of contact interactions, the long-wavelength excitations are not phononlike and always unstable. The post-roton-instability dynamics, in contrast to phonon instability, in a uniform condensate, is featured by a transient, defect-free, stripe pattern, which eventually undergoes local collapses, and driving the condensate back into the stable regime can make them sustained for longer. Hopping between stripes has been observed before it melts into a uniform state in the presence of dissipation. Finally, we discuss a class of solutions, in which a quasi-two-dimensional condensate
is self-trapped in one direction, as well as a regime of interaction parameters, including attractive short-range interactions, at which a two-dimensional anisotropic soliton can be stabilized, and we show that a chromium condensate with a relatively small number of atoms is well suited for this.
\end{abstract}

\pacs{}

\keywords{}

\maketitle


\section {introduction}
Bose-Einstein condensates (BECs) of Chromium (Cr) \cite{crbec_05}, Erbium (Er) \cite{erbec_12} and Dysprosium (Dy) \cite{dybec_11} provided the possibility to study and observe intriguing properties of dipolar quantum gasses \cite{dip_rvw_09, bar12}. Due to the anisotropic nature of dipole-dipole interactions (DDI), the stability/instability properties  crucially depend on the trapping geometry \cite{dip_san_00}. In particular, a dipolar BEC undergoes collapse instability if the attractive element in the DDI dominates the repulsive counterparts \cite{lah08}. Recent experiments \cite{expt_dy15, expt_dy16_1,expt_er16,expt_dy16_2} featured a different outcome for strong dipoles, in which the collapse is suppressed by quantum fluctuations \cite{lima_Qfl_11,wac_fil16, bis_fl_16, wach_pro_16, bal_qf_16}, resulting in the formation of stable self-bound droplets.

The studies on dipolar condensates in the mean-field regime, neglecting quantum fluctuations, are shown to exhibit rich phenomena \cite{dip_rvw_09, bar12}, in particular, roton-maxon spectrum \cite{san_rot03},  self-bound soliton-like solutions \cite{sol-luis,sol-aniso} and structured ground states \cite{shai_rot_07} in pancake type BECs. In addition, in these cases, the orientation of the dipoles with respect to the condensate plane can introduce new features \cite{td_fed_14, bai_til_15,mul_til_vor13, ds_tilt_16}, for instance,  the anisotropic spectrum \cite {bis_ani_sp12} and superfluidity \cite{tic_Asf_11}. Also, the stability analysis of other dipolar systems \cite{zha_til_10, mac_til_14}, by introducing the tilting angle with respect to the normal vector of the two-dimensional (2D) plane, lead to finding different phases, for example, crystal, super-solid and  liquid crystal \cite{wu_lcry_16} ones, including a stripe phase \cite{mac_str_12}.

In this paper, we study the effect of dipolar orientation with respect to the condensate plane on the mean-field properties and stability of dipolar BECs in a pancake confinement. Especially, we look at the ground states, low-lying excitations, and post roton-instability (RI) dynamics. Crucially, the stability/instability regimes as a function of tilting angle $\alpha$ are found to be different for a uniform pancake condensate in the quasi-2D (Q2D) regime and a uniform 2D layer of dipoles, indicating that the extension of the BEC in the transverse direction is critical while characterizing the system. The tilting angle offers the possibility of an anisotropic roton-like Bogoliubov spectrum \cite{bai_til_15} and we show that the roton momentum can be varied from very low magnitude to the order of the inverse-width of the transverse confinement. Then, we show that the roton may arise solely due to the anisotropic and the momentum dependence of DDI, in the absence of short-range contact interactions. The post RI dynamics is characterized by a dislocation-free stripe pattern followed by collapse instability, which is in contrast to the post-PI dynamics, where the latter exhibits dislocation defects.  Finally, we briefly discuss about self-trapped bright solitonic solutions, including a new class, in which a Q2D BEC is self-trapped in one direction, identical to the one-dimensional (1D) bright solitons \cite{sol_expt_02, sol_expt_Nat_02,sol_expt_06}. The in-plane excitations of the these solutions are studied using a variational-Lagrangian formalism. Also, we found a new regime of interaction parameters, including attractive contact interactions, for stable Q2D anisotropic solitons with sufficiently low dipolar strengths.

The paper is structured as follows. In Sec. II we discuss the model and the corresponding non-local Gross-Pitaevskii equations (NLGPEs). In Sec. III, we examine the Bogoliubov excitations of a pancake-like dipolar condensate with a uniform density in the condensate plane using both Q2D and 3D calculations. The post RI dynamics in a pancake-like uniform condensate is discussed in Sec. IV. The different self-trapped solutions in a Q2D condensate; using both numerical and Gaussian variational calculations are shown in Sec. V. Finally, we conclude in Sec. VI.

\section{Model}
\begin{figure}
\vspace{0.cm}
\centering
\includegraphics[width= 1.\columnwidth]{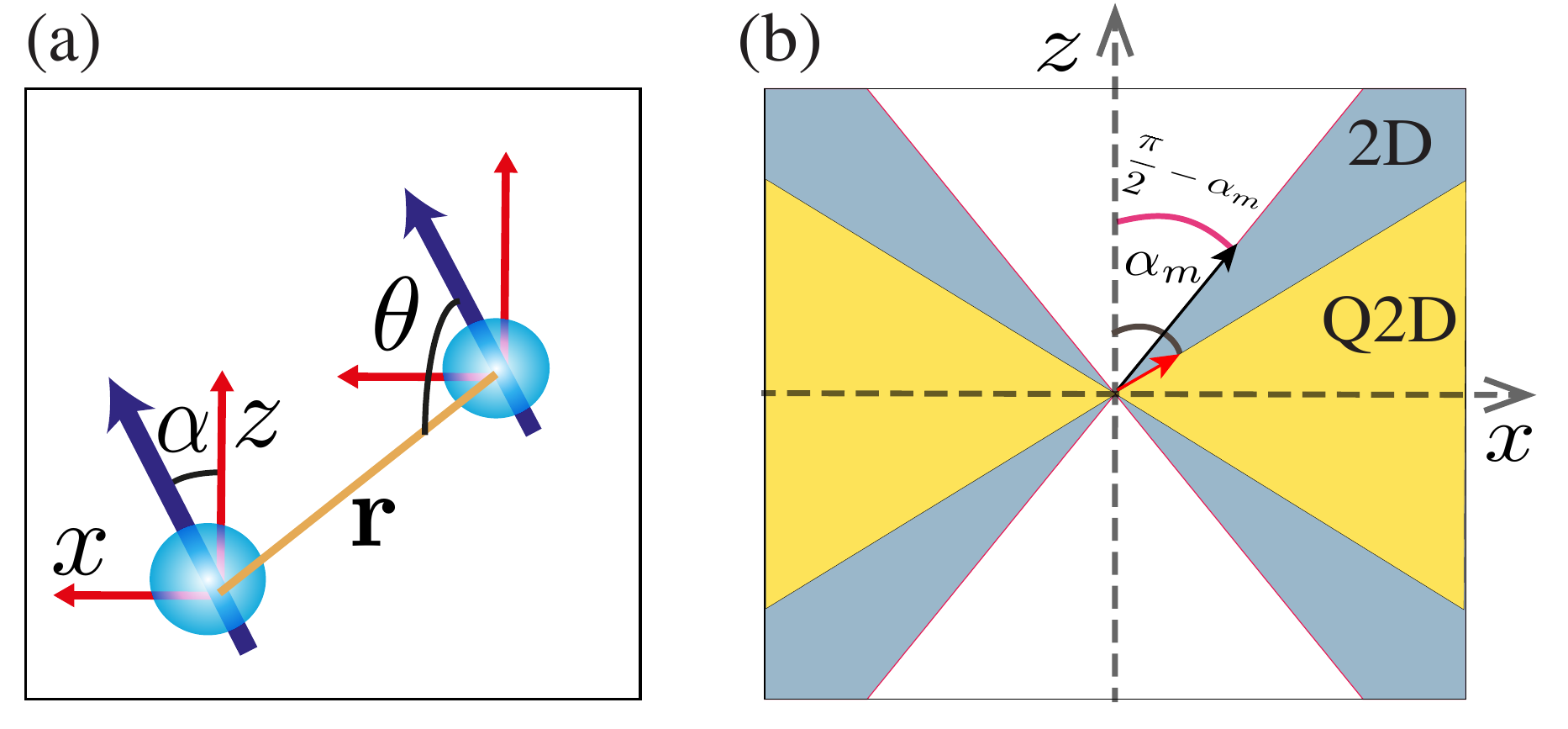}
\caption{\small{(Color online) (a) The dipoles are oriented in the $xz$ plane forming an angle $\alpha$ with the $z$-axis. (b) For $\alpha_m<\alpha< \pi/2$ a purely Q2D dipolar condensate is unstable against collapse, indicated by smaller (yellow) shaded region, but an ideal 2D layer of dipoles is unstable when $\pi/2-\alpha_m<\alpha< \pi/2$, indicated by the larger (gray) shaded region.}}
\label{fig:sd} 
\end{figure}

We consider a BEC of  $N$ particles with magnetic or electric dipole moment $d$, oriented in the $xz$ plane forming an angle $\alpha$ with the $z$-axis, using a sufficiently large external field [Fig. \ref{fig:sd}(a)]. The DDI potential is $V_d({\bf r})=g_d(1-3\cos^2\theta)/r^3$, where $\theta$ is the angle formed by the dipole vector ${\bf d}\equiv d(\sin\alpha \  \hat x + \cos\alpha \ \hat z)$ and the radial vector {\bf r}, with $g_d\propto d^2$ being the strength of the dipole-potential. At low-enough temperatures the system is described by a non-local Gross-Pitaevskii equation (NLGPE):
\begin{equation}
\begin{split}
i\hbar\frac{\partial}{\partial t}\Psi({\bf r},t)=\left[-\frac{\hbar^2}{2m}\nabla^2+V_{t}({\bf r})+g|\Psi({\bf r},t)|^2+\right .\\
\left. \int d{\bf r}^{\prime}V_d({\bf r}-{\bf r}^{\prime})|\Psi({\bf r}^{\prime},t)|^2\right]\Psi({\bf r},t),
\label{gpe3d}
\end{split}
\end{equation}
where $\int d{\bf r}|\Psi({\bf r},t)|^2=N$ and $g=4\pi\hbar^2a/m$ is the coupling constant that characterizes the short-range contact interactions, with $a$ being the $s$-wave scattering length. The trapping potential is $V_t({\bf r})=m(\omega_\rho^2\rho^2+\omega_z^2z^2)/2$ and we assume a pancake type trap hence, the trapping frequencies satisfy: $\omega_{z}>\omega_{\rho}$. For a homogeneous condensate in the $xy$ plane ($\omega_{\rho}=0$), the condensate wave function can be written as $\Psi({\bf r}, t)=\sqrt{n_{2D}}\psi(z) e^{-i\mu t}$ where $n_{2D}$ is the 2D homogeneous density and $\mu$ is the chemical potential. The solution $\psi(z)$ is then described by a local GPE 
\begin{equation}
\left(\frac{-\hbar^2}{2m} \frac{d^2}{dz^2}+\frac{m\omega_z^2 z^2}{2} +g_{eff}n_{2D}|\psi(z)|^2-\mu\right)\psi(z)=0,
\label{gpez}
\end{equation}
with $g_{eff}=g+\frac{4\pi g_d}{3}\left(3\cos^2\alpha-1\right)$ being an effective coupling constant. We have two different regimes based on $\psi(z)$ \cite{par_TF08}: (i) three-dimensional (3D) or Thomas-Fermi (TF) for sufficiently large $g_{eff}$ and (ii) Q2D for small values of $g_{eff}$. Below, we briefly discuss about the Q2D regime.
\subsection{Quasi-2D regime}

 In this regime, we can approximate the transverse wave function $\psi(z)\propto \exp[-z^2/2l_z^2]$, to the ground state of the harmonic oscillator potential along the $z$ axis with a constraint: $|\mu_{2D}|\ll\hbar\omega_z$, where $\mu_{2D}$ is the chemical potential of the Q2D uniform BEC \cite{nath-pho} and $l_z=\sqrt{\hbar/m\omega_z}$. Hence, factorizing the BEC wave function as $\Psi({\bf r})=\psi(x,y)\phi(z)$, then using convolution theorem, the Fourier transform of the DDI potential, 
 \begin {equation}
V_d(k)=\dfrac{4\pi g_d}{3}\left[\dfrac{3(k_x^2\sin^2\alpha+k_x k_z\sin 2\alpha+k_z^2\cos^2\alpha)}{k_x^2+k_y^2+k_z^2}-1\right].
\end {equation}
 and integrating over $dz$, we get an effective 2D NLGPE: 
\begin{equation}
\begin{split}
i\hbar\frac{\partial}{\partial t}\psi(x,y,t)=\left[-\frac{\hbar^2}{2m}\nabla^2_{x,y}+\frac{m\omega_{\rho}^2\rho^2}{2}+\frac{g}{\sqrt{2\pi}l_z}|\psi(x,y,t)|^2+\right .\\
\left. \frac{2 g_d}{3l_z}\int \frac{dk_xdk_y}{(2\pi)^2}e^{i(k_xx+k_yy)}f(k_x,k_y)\ \tilde n(k_x,k_y)\right]\psi(x,y,t),
\label{2dgpe}
\end{split}
\end{equation}
with $\tilde n(k_x,k_y)$ Fourier transform of $|\psi(x,y)|^2$ and
\begin{equation}
\begin{split}
f(k,\theta_k)=\sqrt{2\pi}\left(3\cos^2\alpha-1\right)+3\pi \ e^{k^2/2}k \ {\rm erfc}\left(\frac{k}{\sqrt{2}}\right)\\
\times\left(\sin^2\alpha\cos^2\theta_k-\cos^2\alpha\right),
\end{split}
\end{equation}
where we have used the dimensionless polar coordinates $\big(k\equiv l_z\sqrt{k_x^2+k_y^2}$ and $\theta_k\big)$, and $\mathrm{erfc}(x)$ is the complimentary error function.

  The Q2D homogeneous solution of Eq. \ref{2dgpe} is $\psi(x,y,t)=\sqrt{n_{2D}}\exp[-i\mu_{2D}t/\hbar]$, with  $\mu_{2D}=g_{eff}n_{2D}/\sqrt{2\pi}l_z$. $\mu_{2D}<0$ implies phonon-instability (PI) \cite{mul_vor_til14} and it may lead to either collapse or the formation of a gas of bright solitons \cite{nath-pho}. As noted, a purely dipolar ($g=0$) uniform Q2D BEC is unstable against collapse if $\alpha>\alpha_m=\cos^{-1}\left(1/\sqrt{3}\right)$, interestingly, it is fundamentally different for dipolar bosons in an ideal 2D setup (corresponds to the limit $l_z\to 0$) \cite{mac_str_12} with dipoles occupying $xy$ plane and polarized in the $xz$-plane. In the latter case, the dipole potential is $V_d(x,y)\propto \left(1-3x^2/\rho^2\right)$ with $\rho=\sqrt{x^2+y^2}$ and the instability occurs when $\alpha>\pi/2-\alpha_m$, which is schematically shown in Fig.\ref{fig:sd}(b). In other words, the instability window in $\alpha$ got narrower with a finite $l_z$. This modification of anisotropic behaviour of DDI in $\alpha$ was not so obvious to identify and hence, we stress that the spatial extension of the dipoles in the transverse direction is extremely crucial when characterizing a Q2D system with tilted dipoles, including the case for dipolar fermions \cite{bru_fer_08}.
  
\section{Low lying excitations}

\begin{figure}
\vspace{0.5cm}
\centering
\includegraphics[width= 1.\columnwidth]{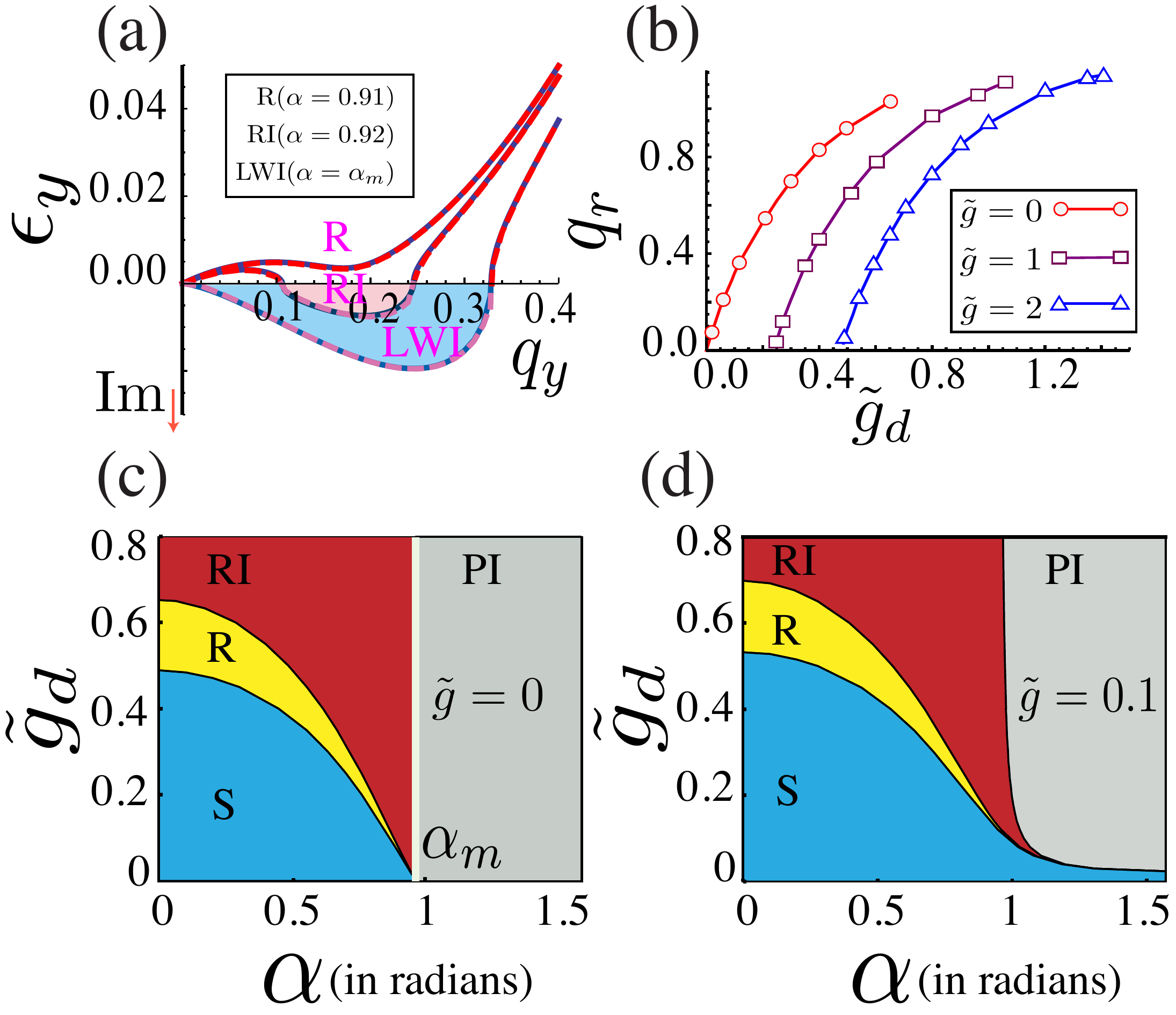}
\caption{\small{(Color online) (a) 3D numerical results for the Bogoliubov excitations along $q_y$ axis, i.e., $\epsilon_y=\epsilon(q_x=0, q_y)$ for different $\alpha$'s for $\tilde g=0$ and $\tilde g_d=0.05$. When $\alpha=\alpha_m$ the condensate is unstable against long wavelength excitations (LWI), $\alpha=0.92$ radians there is RI and for $\alpha=0.91$ radians the spectrum becomes stable and has a maxon-roton behaviour. Since the parameters well satisfy the 2D criteria $\mu_{2D}\ll\hbar\omega_z$, the 3D calculations (dashed red line) are in excellent agreement with the 2D analytical results (solid lines). The imaginary part of $\epsilon_y$ is shown in the negative $y$ axis. (b) The 3D results for the roton momentum $q_r$ as a function of $\tilde g_d$ for different $\tilde g$ values. For a fixed $\tilde g$ and $\tilde g_d$, the value of $\alpha$ is adjusted such that  $\epsilon(q_x=0, q_y=q_r)=0$. The curves are terminated at a value of $g_d$ such that no stable rotons are possible for $0\leq\alpha<\alpha_m$.In (c) and (d) we show the stability/instability regimes of a pancake dipolar condensate with a uniform density in the $xy$ plane, as a function of $\alpha$ and $\tilde g_d$. Figure (c) is for $\tilde g=0$ and (d) for $\tilde g = 0.1$. The abbreviations in the figures (c) and (d) stand for, PI: phonon instability, RI: roton instability, R: stable roton and S: stable spectrum without rotons. In (c), at $\alpha=\alpha_m$, the vertical line indicates the LWI.}}
\label{fig:spd} 
\end{figure}

In this section, we calculate the Bogoliubov dispersion for pancake-like dipolar condensates. They are of the form $\sim f_{\pm}(z)\exp [i{\bf q}\cdot{\rho}-i\epsilon t/\hbar]$, where ${\bf q}$ is the quasi-momentum in the trap-free plane, $\epsilon$ is the excitation energy and $f_{\pm}=u\pm v$ with $\{u, v\}$ are the Bogoliubov functions. The corresponding Bogoliubov-deGennes (BdG) equations, are obtained by linearizing $f_{\pm}(z)$ around $\psi(z)$ and are,
\begin{eqnarray}
\epsilon f_-(z) &=&\left[ \frac{-\hbar^2}{2m}\left(\dfrac{\partial^2}{\partial z^2}-q^2\right)-\mu+\frac{1}{2}m\omega_z^2 z^2+3n_{2D}g_{eff}|\psi(z)|^2\right] f_+(z)\nonumber\\
&& + 4\pi g_dn_{2D} \psi(z)\int_{-\infty}^{\infty}dz' \psi(z') e^{-q_\rho|z'-z|}\left[\frac{q_x^2}{q_\rho}\sin^2\alpha-\right. \nonumber \\
&& \left. q_\rho\cos^2\alpha-i q_x \sin (2\alpha) \rm{sgn}(z'-z)\right]f_+(z')\label{asym}\\
\epsilon f_+(z) &=&\left[ \frac{-\hbar^2}{2m}\left(\dfrac{\partial^2}{\partial z^2}-q^2\right)-\mu + \frac{1}{2}m\omega_z^2 z^2 +n_{2D}g_{eff}|\psi(z)|^2\right] f_-(z). \nonumber \\
\label{bdge}.
\end{eqnarray}
The excitation spectrum $\epsilon({\bf q})$ is obtained by diagonalizing the corresponding Bogoliubov Hamiltonian, and the lowest eigen-value gives us the dispersion in ${\bf q}$. The last term with integral in Eq. \ref{asym} accounts for the momentum dependence of DDI and in particular, leads to the maxon-roton spectrum \cite{san_rot03}. When $\alpha=0$, this term has no influence for excitations with $qL\ll 1$ where $L$ is the spatial width of $\psi(z)$ and we retrieve the linear phonon branch  with a slope $\propto \sqrt{(g+8\pi g_d/3)n_{2D}}$. Interestingly, this is not always the case, for instance, when $\alpha = \alpha_m$ the behaviour of long-wave length (${\bf q}\to 0$) excitations 
are determined by the integral term in Eq. \ref{asym}.  This is easily visible in the Q2D limit ($L=l_z$). In the latter case, the spectrum can be explicitly obtained as \cite{tic_Asf_11,meg15}
\begin{equation}
\epsilon({\bf q})=\sqrt{E_q\left\{E_q+\frac{2gn_{2D}}{\sqrt{2\pi}l_z}\left[1+{\frac{2\sqrt{2\pi}}{3}}\beta f(q,\theta_q)\right]\right\}},
\label{2de}
\end{equation}
and the integral term in Eq. \ref{asym} leads to the the $\alpha$-dependent function $f(q,\theta_q)$ in Eq. \ref{2de}. Now, taking $g=0$ and $\alpha=\alpha_m$, we get  $\epsilon({\bf q}\to 0)\propto \sqrt{g_dq(q_x^2-q_y^2)}$, which are no-longer the phonon-like excitations. These long-wavelength excitations with momenta $q_y>q_x$ are unstable (imaginary) for $g_d>0$. We term this instability as long-wavelength instability (LWI) instead of PI since the chemical potential $\mu_{2D}(g=0,\alpha=\alpha_m)=0$ as well as the excitations are no-longer linear in momenta.  Making $g_d$ larger extends the instability regions to higher momenta and reducing $\alpha$ stabilizes the LWI, together we have a finite-momenta instability or RI and is pre-dominantly along $q_y$ axis. Reducing $\alpha$ further gives us a stable anisotropic spectrum with maxon-roton behaviour along the $q_y$ axis. The Q2D results (Eq. \ref{2de}) for the excitations [$\epsilon_y=\epsilon(q_x=0, q_y)$] are shown in Fig. \ref{fig:spd}(a), which are in excellent agreement with the 3D results obtained via Eqs. \ref{asym} and \ref{bdge}. The imaginary part of $\epsilon_y$ is shown in the negative $y$ axis. Note that, for convenience we introduced the dimensionless interaction variables $\tilde g=gn_{2D}/\hbar\omega_zl_z$ and $\tilde g_d=g_dn_{2D}/\hbar\omega_zl_z$.

\begin{figure*}[hbt]
\centering
\includegraphics[width= .95\textwidth]{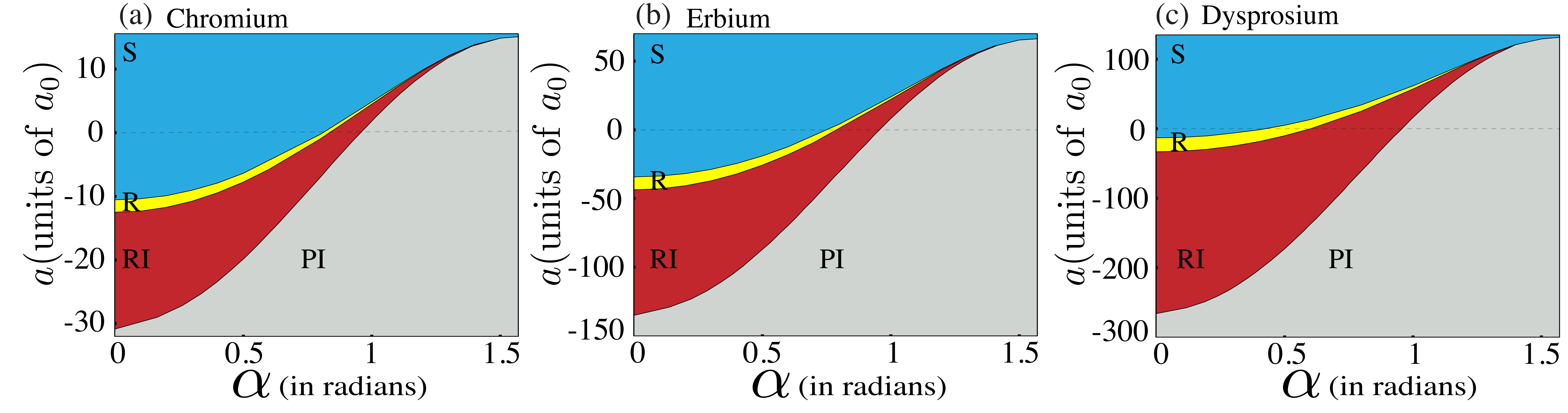}
\caption{\small{(Color online) The stability/instability regions of the pancake-like dipolar condensates with a uniform density in the $xy$ plane, as a function of the tilting angle $\alpha$ and the $s$-wave scattering length $a$ (in units of Bohr radius $a_0$) for Cr, Er and Dy BECs, with $\omega_z=2\pi\times 800$Hz and the condensate density $n_{2D}/\left(\sqrt{2\pi}l_z\right)=10^{20}m^{-3}$, based on the Bogoliubov modes obtained by solving Eqs. \ref{asym} and \ref{bdge}. The dipolar coupling constant $\tilde g_d$ is calculated using the intrinsic magnetic moments of each of the atoms. The abbreviations in the figure stand for, PI: phonon instability, RI: roton instability, R: stable roton and S: stable spectrum without rotons.}}
\label{fig:atms} 
\end{figure*}

There are two very interesting perspectives associated with the anisotropic rotons in Q2D BECs. First, unlike the isotropic rotons ($\alpha=0$) \cite{nath_far_10}, which requires $g<0$, here the anisotropic ones arise entirely ($g=0$) from the anisotropic nature and the momentum dependence of DDI. The second and the most important is, on contrary to previous studies where the roton momentum at which $\epsilon(q=q_r)=0$ is found to be, $q_r\sim 1/l_z$, given by the inverse of the transverse width, here,  in this case, $q_r$ can be varied continuously from very low values to the order of $1/l_z$, by changing the interaction parameters. As an example, in Fig. \ref{fig:spd}(b), we show $q_r$ (3D results) as a function of $g_d$ for different $g$ values. The value of $\alpha$ is chosen such that $\epsilon(q_x=0, q_y=q_r)=0$. The curves are terminated at a maximum value of $\tilde g_d$ beyond which no stable rotons are possible for $0\leq\alpha<\pi/2$.  Note that, when $\alpha=0$, we have a maxon-roton spectrum with $g=0$, but it is for large values of $g_d$ such that the Q2D criteria doesn't hold \cite{uwe_dip_06}.

At this point, we summarize all our above observations as diagrams exhibiting stability/instability regimes of the pancake condensate as a function of $\alpha$ and $\tilde g_d$. The results are shown in Fig. \ref{fig:spd}(c) and (d) respectively for $\tilde g=0$ and $\tilde g\neq 0$. In the absence of contact interactions ($\tilde g=0$), PI occurs for $\alpha>\alpha_m$ assuming $g_d>0$. With a finite $\tilde g(>0)$, the stable region (S) gets extended beyond $\alpha_m$ for lower values of $\tilde g_d$, see Fig. \ref{fig:spd}(d), until the attractive part of the DDI dominates the effective repulsive interactions leading to PI. In the end, we provide the stability regimes for the state of the art experimental setups, such as Cr, Er and Dy BECs, see Fig. \ref{fig:atms}, as a function of $a$ and $\alpha$. The stability of a Cr condensate  as a function of aspect ratio $\lambda=\omega_z/\omega_{\rho}$ and $a$, for $\alpha=0$, has been probed experimentally up to a maximum value of $\lambda=10$ \cite{koc_nat_08}. The experimental predictions are found to be in excellent agreement with the calculations based on a Gaussian density profile and in the limit $\lambda\to \infty$ the collapse instability is predicted to be for $a\sim -30a_0$, which agrees very well with our results as well.

\section{Post instability dynamics}
\begin{figure}
\vspace{0.5cm}
\centering
\includegraphics[width= 1.\columnwidth]{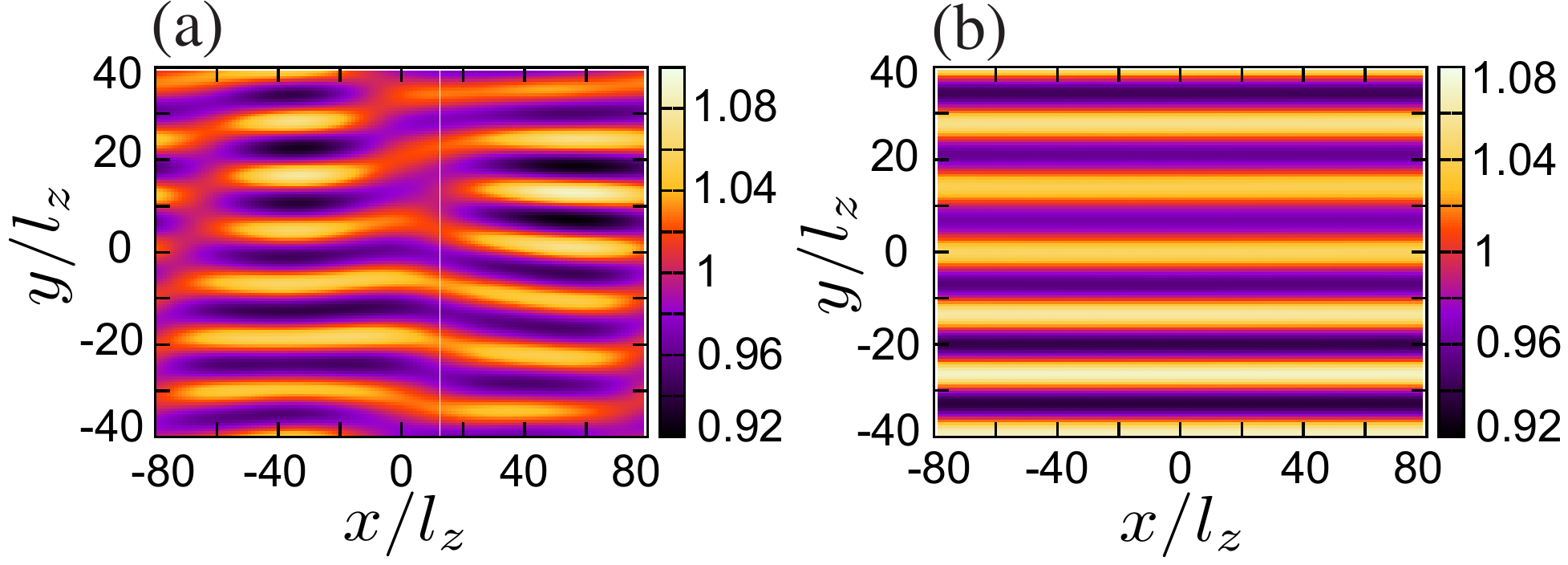}
\caption{\small{(Color online) (a) The post-PI transient stripe pattern with dislocation defects in a Q2D homogeneous condensate and (b) the stripe pattern without defects in post-RI dynamics. The roton softening happens along the $q_y$ axis and hence the stripes are parallel to $x$ axis. The formation of defect-free stripes can be identified as the trait of anisotropic RI in dipolar BECs. For the simulation we have considered a Cr BEC with density $n_{2D}/\sqrt{2\pi}l_z=5\times 10^{19}m^{-6}$, $\omega_z=2\pi\times 800$ Hz and for (a)  $a=0$ and $\alpha=\alpha_m+0.14$ radians and (b)  $a=5a_0$ and $\alpha=\alpha_m-0.14$ radians with $a_0$ being the Bohr radius. The snapshots are taken at (a) $t=0.1s$ and (b) $t=0.016 s$.}}
\label{fig:pr} 
\end{figure}

As shown in \cite{meg15}, the post-PI dynamics with $\alpha\neq 0$ shows a transient stripe pattern that present dislocation defects and eventually the stripes breaks up into anisotropic solitons. At the dislocation defects each stripes merge into one. The formed solitons fuse together to form a bigger one, and may get unstable against collapse if its density goes beyond a critical value. Here, we discuss the post-RI dynamics in a pancake-like dipolar condensate with a uniform density in the $xy$ plane. Starting from a stable (S) regime [see Fig. \ref{fig:atms}] the RI can be attained by many ways, for instance, varying the scattering length $a$ or changing the tilting angle $\alpha$. Interestingly, on contrary to the PI case, the post-RI dynamics is characterized by a defect-free stripe pattern and eventually they become unstable against local collapses. Hence, the dislocation free stripes pattern can be identified as a signature of (anisotropic) roton-softening in dipolar condensates. The patterns for both PI and RI are shown in Fig. \ref{fig:pr}. In the case of RI, the unstable roton momenta are localized along the $q_y$ axis hence, the stripes are parallel to the $x$-axis, whereas for PI with $\alpha_m\leq\alpha\leq\pi/2$ there is always a small component of instability momenta along $q_x$ axis resulting in the dislocation of stripes. The stripes formed via RI can be made to sustain for longer period of time by driving the system slowly back to the roton-stable regime by changing the interaction parameters. If dissipation is present, the amplitude of stripes decay in time and finally the condensate evolves into the uniform state. This has been verified using both 3D and 2D numerical calculations and we have also observed the hopping between stripes in real time, indicating a metastable stripe super-solid-like state.

\vspace{0.5cm}
 \section{Self-trapping in Q2D condensates}
\label{st}
\begin{figure}
\vspace{0.5cm}
\centering
\includegraphics[width= 1.\columnwidth]{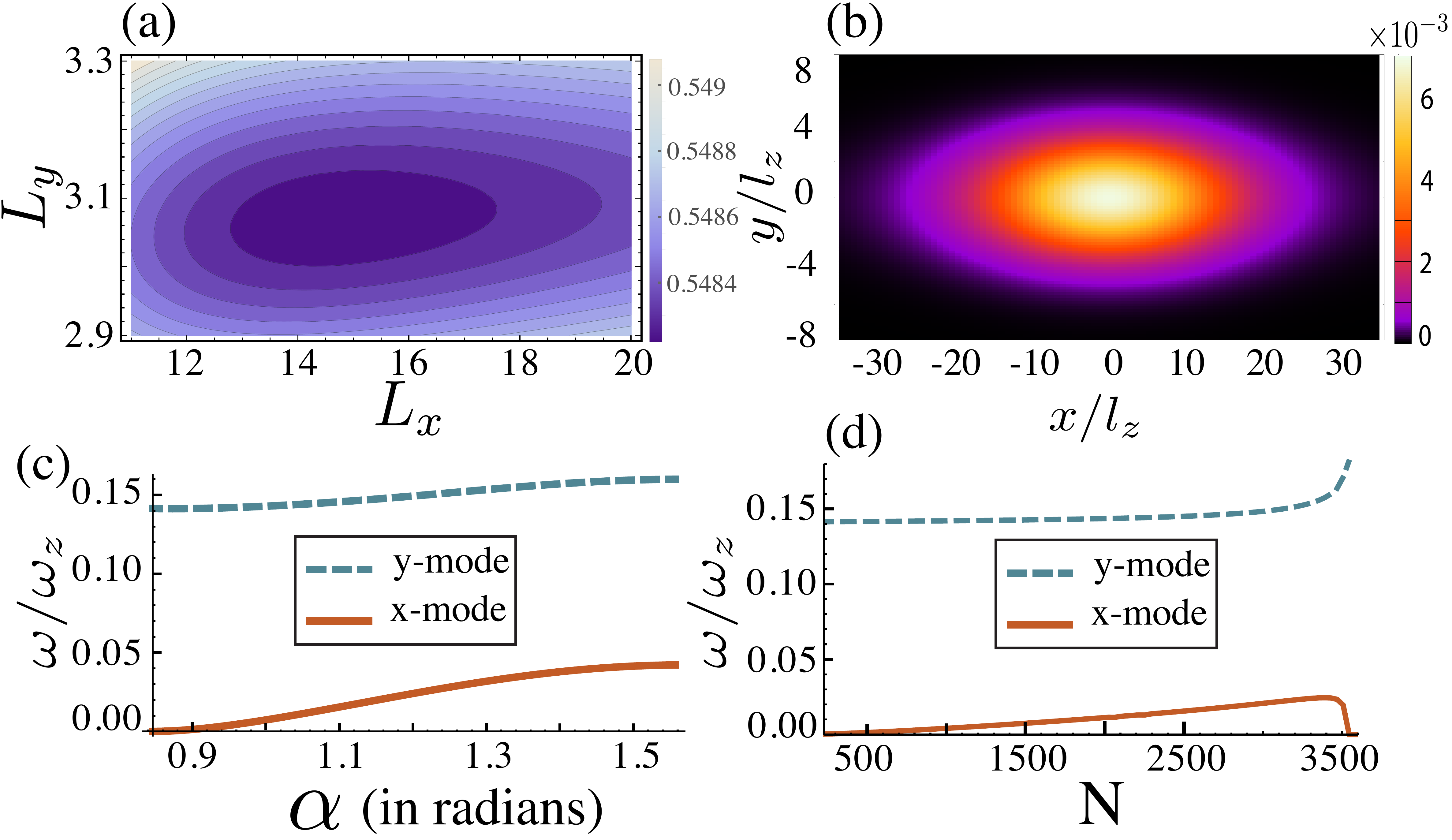}
\caption{\small{(Color online) (a) The contour plot for the energy $E(L_x,L_y,L_z)$ as a function of $L_x$ and $L_y$ with $L_z=1$ for a Chromium condensate with number of particles $N=1000$, $\alpha=\alpha_m$ and $a=0$. The trapping frequencies are taken as $\omega_z=2\pi\times 800$Hz and $\omega_y=2\pi\times 80$Hz. The self-trapping along $x$ axis is evident from the plot. (b) The corresponding ground state density $|\psi(x,y)|^2$ in the $xy$ plane. (c) The low-lying $xy$-modes as a function of the tilting angle ($\alpha$) and other parameters are same as that of (a) and (b). (d) The low-lying $xy$-modes as a function of the number of atoms in the condensate ($N$) and other parameters are same as that of (a) and (b) }}
\label{fig:ep} 
\end{figure}

Having discussed the stability properties of dipolar condensates in a pancake-confinement, we analyze the self-trapped bright solitonic solutions in this setup. As we know, in general, the bright solitons are stable in quasi-1D (Q1D) geometries \cite{sol_expt_02, sol_expt_Nat_02,sol_expt_06}, while in 2D and 3D they are unstable against collapse. DDI together with short-range interactions ($g>0$) may stabilize 2D bright solitons including both isotropic  \cite{sol-luis} and anisotropic ones \cite{sol-aniso,meg15}. Below, we show two results:  (i)  a new kind of self-trapped solutions in which a Q2D dipolar BEC is self-confined in one direction, identical to that of a 1D bright soliton and (ii) a new regime of interaction parameters for stable Q2D anisotropic solitons, with $g<0$ and sufficiently low effective dipolar strengths. In the former case, due to the Q2D nature of the condensate, the self-trapping in one direction cannot be purely described by a 1D NLGPE and hence, strictly not a 1D soliton. In order to see these properties, we introduce the condensate energy functional
\begin{equation}
\begin{split}
E=\int d^3r\left[\frac{\hbar^2}{2m}|\nabla\Psi_0(r)|^2+V_t(y,z)|\Psi_0(r)|^2+\frac{g}{2}|\Psi_0(r)|^4+\right. \nonumber \\
\left.\frac{1}{2}\int d^3r^{\prime}V_d(r-r^{\prime})|\Psi_0(r)|^2|\Psi_0(r^{\prime})|^2\right],
\end{split}
\end{equation}
with harmonic confinements along both $y$ and $z$ directions, and using a Gaussian ansatz of the form:
\begin{equation}
\Psi_0({\bf r})=\frac{1}{\pi^{3/4}l_z^{3/2}\sqrt{L_xL_yL_z}}\exp\left[-\frac{1}{2l_z^2}\left(\frac{x^2}{L_x^2}+\frac{y^2}{L_y^2}+\frac{z^2}{L_z^2}\right)\right],
\label{gau}
\end{equation}
we obtain
\begin{widetext}
\begin{equation}
\begin{split}
\frac{E}{\hbar\omega_z}=\frac{1}{4L_x^2}+\frac{1}{4L_y^2}+\frac{1}{4L_z^2}+\frac{L_z^2}{4}+\frac{\lambda^2 L_y^2}{4}+\frac{\bar g}{4\pi L_xL_yL_z}+\frac{\bar{g}_d}{3L_z}\left[\frac{3\sin^2\alpha}{L_x^2-L_y^2}\left(\sqrt{\frac{L_y^2-L_z^2}{L_x^2-L_z^2}}-\frac{L_y}{L_x}\right)+\frac{3\cos^2\alpha}{\sqrt{\left(L_x^2-L_z^2\right)\left(L_y^2-L_z^2\right)}}-\frac{1}{L_xL_y}\right] \\
-\frac{2\bar g_d}{\pi}\int_0^{\pi/2}d\chi\frac{\cos^2\chi\sin^2\alpha-\cos^2\alpha}{\left[L_z^2-\left(L_x^2\cos^2\chi+L_y^2\sin^2\chi\right)\right]^{3/2}} \ \  {\rm arctanh}\left(\frac{\sqrt{L_z^2-\left(L_x^2\cos^2\chi+L_y^2\sin^2\chi\right)}}{L_z}\right),
\end{split}
\label{E3d}
\end{equation}
\end{widetext}
where $\lambda=\omega_y/\omega_z$, $\bar g=gN/\sqrt{2\pi}\hbar\omega_zl_z^3$ and $\bar g_d=g_dN/\sqrt{2\pi}\hbar\omega_zl_z^3$. Note that, first we look for self-trapping along the $x$ axis. The minimum [see Fig. \ref{fig:ep}(a)] of $E$ [$E^{\rm{min}}=E(L_x^{\rm{min}}, L_y^{\rm{min}}, L_z^{\rm{min}})$] provides us the equilibrium widths, $w_i^0=l_zL_i^{\rm{min}}$ with $i\in\{x, y, z\}$, of the condensate density. The numerical solution using 2D NLGPE with the same  parameters of Fig. \ref{fig:ep}(a)  is shown in Fig. \ref{fig:ep}(b). The absence of a minimum in $E$ can arise from two distinct types of instabilities. If the repulsive part of the interactions dominates, the soliton expands without limits in the $x$ direction ($w_{x}^0\to\infty$), whereas dominating attractive interactions lead to collapse ($w_{x,y,z}^0\to 0$). If $\lambda=1$ and $\hbar\omega_{z}\gg\mu_{1D}$ (criteria for Q1D regime) where $\mu_{1D}$ is the chemical potential of the Q1D condensate, the collapse can be suppressed completely and we get a 1D bright soliton solution along the $x$ axis.

\begin{figure}
\vspace{0.5cm}
\centering
\includegraphics[width= 1.\columnwidth]{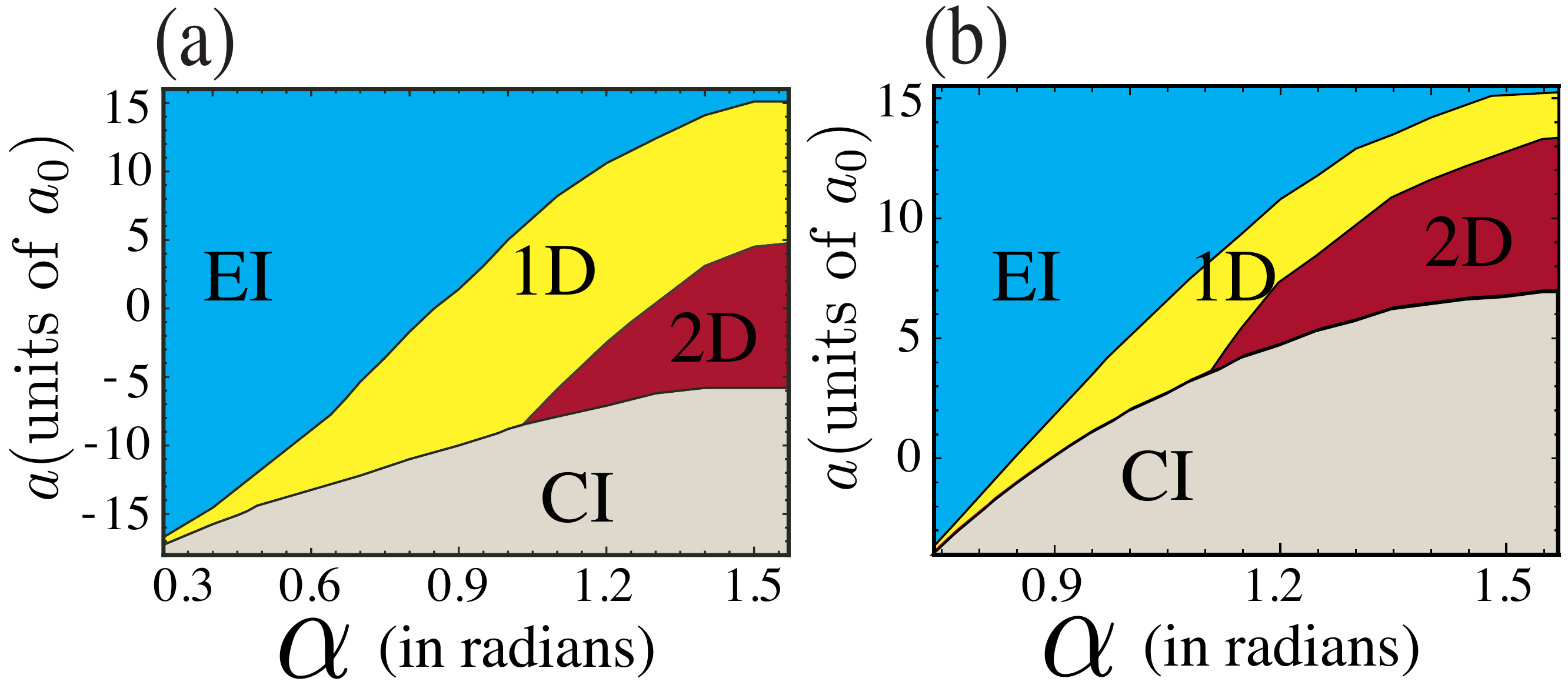}
\caption{\small{(Color online) The stability/instability regimes of self-trapped solutions in a Q2D dipolar condensate as a function of $a$ and $\alpha$. We consider a Cr condensate with $\omega_z=2\pi\times 800$Hz, $\omega_y=2\pi\times 80$Hz and $N=1000$ for (a) and $N=5000$ for (b). The abbreviations in the figure stand for, EI: expansion instability, CI: collapse instability, 1D: 1D self-trapping and 2D: 2D anisotropic soliton.}}
\label{fig:pd:st} 
\end{figure}

Once the equilibrium widths are found, we examine the lowest-lying modes ($\omega$) of these self-trapped solutions using a variational method \cite{var_zoll_96,var_you_01}, where we use a time-dependent Gaussian as the trial wave function: 
\begin{eqnarray}
\psi(x,y,z,t) &=& A(t)\prod\limits_{\eta=x,y,z} e^{-\frac{(\eta-\eta_0(t))^2}{2w_\eta^2(t)}} e^{i\eta\alpha_\eta(t)} e^{i\eta^2\beta_\eta(t)},
\end{eqnarray}
where $\eta_0$, $w_\eta$, $\alpha_\eta$ and $\beta_\eta$ are the time dependent variational parameters and the normalization constant $A(t)=\pi^{-3/4}/\sqrt{w_xw_yw_z}$. The above ansatz is then introduced in the Lagrangian density of a dipolar BEC:
\begin{widetext}
\begin{eqnarray}
\mathcal L&=& \frac{i}{2}\hbar\left( \psi\frac{\partial{\psi^*}}{\partial{t}}-\psi^*\frac{\partial{\psi}}{\partial{t}}\right)+\frac{\hbar^2}{2m}|\nabla\psi(r,t)|^2 + V_t(y,z)|\psi(r,t)|^2+\frac{g}{2}|\psi(r,t)|^2+\frac{1}{2}|\psi(r,t)|^2 \int dr' V_d(r-r')|\psi(r',t)|^2.
\label{lad}
\end{eqnarray}
\end{widetext}
The  Lagrangian is then obtained by integrating over the whole space i.e., $L=\int d^3r\mathcal L$. We then obtain the corresponding Euler-Lagrange equations of motion for the time-dependent variational parameters, see App. \ref{lang}. We consider the solution is static along the $x$ axis and take $x_0(t)=0$, hence $\alpha_x(t)=0$.  Note that the centre of mass motion (along the $z$ and $y$ axis) is decoupled from the internal dynamics of the condensate. We analyze the two in-plane ($xy$-) modes of the condensate as a function $\alpha$ and $N$, which lie lowest in the excitation spectrum. Due to the high anisotropic structure of the condensate in the $xy$ plane, the modes decouple into pure $x$ and $y$ modes. Since it is more elongated along the $x$-axis, the lowest mode corresponds to the oscillation of the condensate width along the $x$-axis. The modes as a function of $\alpha$ is shown in  Fig. \ref{fig:ep}(c) for Cr atoms with $a=0$, $N=1000$, $\omega_z=2\pi\times 800$Hz and $\omega_y=2\pi\times 80$Hz. When $\alpha$ decreases, the lowest mode gets softer, becomes zero below $\alpha_c$ indicating the expansion instability along the $x$ axis. It is because, as $\alpha$ decreases the effective attractive interaction along the $x$-axis decreases. The critical angle, $\alpha_c$ depends only on the ratio $\bar g_d/\bar g$ hence, independent of $N$. Next, we obtain the modes as a function of $N$, see Fig. \ref{fig:ep}(d). Mode softening can be seen at both low and high $N$ values, indicating the two distinct instabilities. For low $N$, the interactions are weaker and the self-trapping is lost against the expansion instability. Conversely,  at high $N$ it undergoes collapse instability. We have also noticed that a similar setup discussed in Fig. \ref{fig:ep}(a) replacing Cr with Er and Dy atoms, leads to collapse of the condensate due to their large dipole moments, but for those systems the quantum fluctuations may become prominent and the scenario can also be different \cite{lima_Qfl_11,wac_fil16, bis_fl_16, wach_pro_16, bal_qf_16}. Therefore, here, we focus on Cr atoms for which quantum fluctuations may be neglected due to its relatively low dipole moment. Finally, we provide stability diagrams for a Cr condensate for $N=1000$ [Fig. \ref{fig:pd:st}(a)] and $N=5000$ [Fig. \ref{fig:pd:st}(b)]. As the results show, the 1D self-trapping in a Q2D condensate is a dominant feature emerges at systems with relatively smaller effective dipolar strengths ($\bar g_d$), which can be attained either with small $N$ or atoms with small dipole moments. As $N$ increases (or equivalently systems with large dipole moments), the stability regime for 1D self-trapping gets smaller and smaller. For $\alpha$ close to $\pi/2$, the 1D self-trapping is separated from collapse instability through 2D anisotropic solitons ($\omega_y=0$). A surprising result that emerged from these studies, is the possibility to stabilize 2D solitons for negative scattering lengths at low $N$ [Fig. \ref{fig:pd:st}(a)]. Hence, we point out that, the fugitive anistropic solitons in the mean-field regime are easily accessible in systems with lower dipolar strengths than higher ones, with sufficiently small number of atoms.

\section{conclusion}
In conclusion, we studied the physics of dipolar condensates in a pancake confinement as a function of the orientation of the dipoles with respect to the condensate plane. The stability regime of a quasi-two-dimensional condensate as a function of tilting angle is found to be different from a two-dimensional layer of dipoles. An anisotropic roton like spectrum may arise entirely due to the DDI and the post RI dynamics in a uniform pancake-like condensate is featured by a transient, defect-free, stripe pattern, which eventually breaks into local collapses. The pattern can be made to sustain for longer time by driving the system back into the stable regime. Hopping between the stripes is observed before it melts into a uniform state in the presence of dissipation. A new class of self-trapped solutions are found, in which a Q2D condensate is self-trapped in one direction, identical to the case of a 1D bright soliton. In the end, a new regime of interaction parameters, with attractive contact interactions for 2D anisotropic solitons are discussed.

\section{Acknowledgments}
We thank Luis Santos and Paolo Pedri for useful discussions and R. N. acknowledges funding by the Indo-French
Centre for the Promotion of Advanced Research - CEFIPRA. C. M. acknowledges the funding from DST India through INSPIRE
fellowship.
\appendix
\section{Lagrangian for a dipolar condensate}
\label{lang}
The Lagrangian density for a dipolar condensate is given in Eq. \ref{lad}, and using the Gaussian time-dependent trial function we obtain the Lagrangian $L=\int d^3r\mathcal L$:
\begin{eqnarray}
L &=& \sum_{\eta=x,y,z}\left[\hbar\frac{\dot{\beta}_\eta w_\eta^2}{2}+\frac{\hbar^2}{2m} \left( \frac{1}{2w_\eta^2}+ \alpha_\eta^2 + 2\beta_\eta^2 w_\eta^2 \right)\right] +\frac{1}{2}m\omega_z^2 \frac{w_z^2}{2}\nonumber \\
&&+\frac{1}{2}m\omega_y^2 \frac{w_y^2}{2}+ \frac{g}{\sqrt{2\pi}}\frac{1}{4\pi w_x w_yw_z} + V(w_\eta), \nonumber \\
\label{lagrangian}
\end{eqnarray}
where 
\begin{equation}
 V(w_\eta)= \frac{1}{2}\frac{1}{\left(2\pi\right)^3} \int d^3k \tilde V_d({\bf k}) \prod\limits_{\eta=x,y,z}e^{-\frac{k_\eta^2 w_\eta^2}{2}}
\end{equation}
with $\tilde V_d({\bf k})$ the Fourier transform the dipole-dipole potential. Then, the equations of motion for $\alpha(t)$ and $\beta(t)$ are:
\begin{eqnarray}
\alpha_{\eta}= \frac{m}{\hbar}\left(\dot\eta_0-\frac{\dot w_{\eta}\eta_0}{w_{\eta}}\right)\\
\beta_{\eta}=\frac{m\dot w_{\eta}}{2\hbar w_\eta} 
\end{eqnarray}
with $\eta\in\{x,y,z\}$ and the same for the condensate widths are
\begin{eqnarray} 
m\ddot w_x &=& \frac{\hbar^2}{mw_x^3} + \frac{g}{(2\pi)^{3/2} w_x^2w_yw_z} - 2\frac{\partial{V}}{\partial{w_x}} \\
m\ddot w_y &=& \frac{\hbar^2}{mw_y^3} + \frac{g}{(2\pi)^{3/2} w_xw_y^2w_z}  -m\omega_y^2y^2-2\frac{\partial{V}}{\partial{w_y}} \\
m\ddot w_z &=& \frac{\hbar^2}{mw_z^3} + \frac{g}{(2\pi)^{3/2} w_xw_yw_z^2} -m\omega_z^2z^2- 2\frac{\partial{V}}{\partial{w_z}}.
\end{eqnarray}
The above equations describe the motion of a particle with coordinates $w_\eta$ in an effective potential
\begin{eqnarray}
U(W_\eta) &=&  \frac{\hbar^2}{2m}\sum_\eta\frac{1}{w_\eta^2}+ \frac{1}{2}m\omega_z^2w_z^2+ \frac{g}{(2\pi)^{3/2} w_xw_yw_z} + V(w_\eta).\nonumber \\
\end{eqnarray}
Once the equilibrium widths of the condensate are obtained by minimizing the effective potential (or equivalently from the Gaussian energy calculations in Section. \ref{st}), the low lying excitations  are obtained  by diagonalizing the Hessian matrix of $U$. Also, note that the centre of mass motion of the soliton along the $z$ axis is de-coupled from the internal dynamics and is governed by the equation
\begin{equation}
\ddot z_0=-\omega_z^2z_0.
\end{equation}
and
\begin{equation}
\ddot y_0=-\omega_y^2y_0.
\end{equation}

 \bibliographystyle{apsrev}
\bibliography{liball1.bib}

\begin{thebibliography}{43}
\expandafter\ifx\csname natexlab\endcsname\relax\def\natexlab#1{#1}\fi
\expandafter\ifx\csname bibnamefont\endcsname\relax
  \def\bibnamefont#1{#1}\fi
\expandafter\ifx\csname bibfnamefont\endcsname\relax
  \def\bibfnamefont#1{#1}\fi
\expandafter\ifx\csname citenamefont\endcsname\relax
  \def\citenamefont#1{#1}\fi
\expandafter\ifx\csname url\endcsname\relax
  \def\url#1{\texttt{#1}}\fi
\expandafter\ifx\csname urlprefix\endcsname\relax\def\urlprefix{URL }\fi
\providecommand{\bibinfo}[2]{#2}
\providecommand{\eprint}[2][]{\url{#2}}

\bibitem[{\citenamefont{Griesmaier et~al.}(2005)\citenamefont{Griesmaier,
  Werner, Hensler, Stuhler, and Pfau}}]{crbec_05}
\bibinfo{author}{\bibfnamefont{A.}~\bibnamefont{Griesmaier}},
  \bibinfo{author}{\bibfnamefont{J.}~\bibnamefont{Werner}},
  \bibinfo{author}{\bibfnamefont{S.}~\bibnamefont{Hensler}},
  \bibinfo{author}{\bibfnamefont{J.}~\bibnamefont{Stuhler}}, \bibnamefont{and}
  \bibinfo{author}{\bibfnamefont{T.}~\bibnamefont{Pfau}},
  \bibinfo{journal}{Phys. Rev. Lett.} \textbf{\bibinfo{volume}{94}},
  \bibinfo{pages}{160401} (\bibinfo{year}{2005}).

\bibitem[{\citenamefont{Aikawa et~al.}(2012)\citenamefont{Aikawa, Frisch, Mark,
  Baier, Rietzler, Grimm, and Ferlaino}}]{erbec_12}
\bibinfo{author}{\bibfnamefont{K.}~\bibnamefont{Aikawa}},
  \bibinfo{author}{\bibfnamefont{A.}~\bibnamefont{Frisch}},
  \bibinfo{author}{\bibfnamefont{M.}~\bibnamefont{Mark}},
  \bibinfo{author}{\bibfnamefont{S.}~\bibnamefont{Baier}},
  \bibinfo{author}{\bibfnamefont{A.}~\bibnamefont{Rietzler}},
  \bibinfo{author}{\bibfnamefont{R.}~\bibnamefont{Grimm}}, \bibnamefont{and}
  \bibinfo{author}{\bibfnamefont{F.}~\bibnamefont{Ferlaino}},
  \bibinfo{journal}{Phys. Rev. Lett.} \textbf{\bibinfo{volume}{108}},
  \bibinfo{pages}{210401} (\bibinfo{year}{2012}).

\bibitem[{\citenamefont{Lu et~al.}(2011)\citenamefont{Lu, Burdick, Youn, and
  Lev}}]{dybec_11}
\bibinfo{author}{\bibfnamefont{M.}~\bibnamefont{Lu}},
  \bibinfo{author}{\bibfnamefont{N.~Q.} \bibnamefont{Burdick}},
  \bibinfo{author}{\bibfnamefont{S.~H.} \bibnamefont{Youn}}, \bibnamefont{and}
  \bibinfo{author}{\bibfnamefont{B.~L.} \bibnamefont{Lev}},
  \bibinfo{journal}{Phys. Rev. Lett.} \textbf{\bibinfo{volume}{107}},
  \bibinfo{pages}{190401} (\bibinfo{year}{2011}).

\bibitem[{\citenamefont{Lahaye et~al.}(2009)\citenamefont{Lahaye, Menotti,
  Santos, Lewenstein, and Pfau}}]{dip_rvw_09}
\bibinfo{author}{\bibfnamefont{T.}~\bibnamefont{Lahaye}},
  \bibinfo{author}{\bibfnamefont{C.}~\bibnamefont{Menotti}},
  \bibinfo{author}{\bibfnamefont{L.}~\bibnamefont{Santos}},
  \bibinfo{author}{\bibfnamefont{M.}~\bibnamefont{Lewenstein}},
  \bibnamefont{and} \bibinfo{author}{\bibfnamefont{T.}~\bibnamefont{Pfau}},
  \bibinfo{journal}{Reports on Progress in Physics}
  \textbf{\bibinfo{volume}{72}}, \bibinfo{pages}{126401}
  (\bibinfo{year}{2009}).

\bibitem[{\citenamefont{Baranov et~al.}(2012)\citenamefont{Baranov, Dalmonte,
  Pupillo, and Zoller}}]{bar12}
\bibinfo{author}{\bibfnamefont{M.~A.} \bibnamefont{Baranov}},
  \bibinfo{author}{\bibfnamefont{M.}~\bibnamefont{Dalmonte}},
  \bibinfo{author}{\bibfnamefont{G.}~\bibnamefont{Pupillo}}, \bibnamefont{and}
  \bibinfo{author}{\bibfnamefont{P.}~\bibnamefont{Zoller}},
  \bibinfo{journal}{Chem. Rev.} \textbf{\bibinfo{volume}{112}},
  \bibinfo{pages}{5012} (\bibinfo{year}{2012}).

\bibitem[{\citenamefont{Santos et~al.}(2000)\citenamefont{Santos, Shlyapnikov,
  Zoller, and Lewenstein}}]{dip_san_00}
\bibinfo{author}{\bibfnamefont{L.}~\bibnamefont{Santos}},
  \bibinfo{author}{\bibfnamefont{G.~V.} \bibnamefont{Shlyapnikov}},
  \bibinfo{author}{\bibfnamefont{P.}~\bibnamefont{Zoller}}, \bibnamefont{and}
  \bibinfo{author}{\bibfnamefont{M.}~\bibnamefont{Lewenstein}},
  \bibinfo{journal}{Phys. Rev. Lett.} \textbf{\bibinfo{volume}{85}},
  \bibinfo{pages}{1791} (\bibinfo{year}{2000}).

\bibitem[{\citenamefont{Lahaye et~al.}(2008)\citenamefont{Lahaye, Metz,
  Fr\"ohlich, Koch, Meister, Griesmaier, Pfau, Saito, Kawaguchi, and
  Ueda}}]{lah08}
\bibinfo{author}{\bibfnamefont{T.}~\bibnamefont{Lahaye}},
  \bibinfo{author}{\bibfnamefont{J.}~\bibnamefont{Metz}},
  \bibinfo{author}{\bibfnamefont{B.}~\bibnamefont{Fr\"ohlich}},
  \bibinfo{author}{\bibfnamefont{T.}~\bibnamefont{Koch}},
  \bibinfo{author}{\bibfnamefont{M.}~\bibnamefont{Meister}},
  \bibinfo{author}{\bibfnamefont{A.}~\bibnamefont{Griesmaier}},
  \bibinfo{author}{\bibfnamefont{T.}~\bibnamefont{Pfau}},
  \bibinfo{author}{\bibfnamefont{H.}~\bibnamefont{Saito}},
  \bibinfo{author}{\bibfnamefont{Y.}~\bibnamefont{Kawaguchi}},
  \bibnamefont{and} \bibinfo{author}{\bibfnamefont{M.}~\bibnamefont{Ueda}},
  \bibinfo{journal}{Phys. Rev. Lett.} \textbf{\bibinfo{volume}{101}},
  \bibinfo{pages}{080401} (\bibinfo{year}{2008}).

\bibitem[{\citenamefont{Kadau et~al.}(2016)\citenamefont{Kadau, Schmitt,
  Wenzel, Wink, Maier, Ferrier-Barbut, and Pfau}}]{expt_dy15}
\bibinfo{author}{\bibfnamefont{H.}~\bibnamefont{Kadau}},
  \bibinfo{author}{\bibfnamefont{M.}~\bibnamefont{Schmitt}},
  \bibinfo{author}{\bibfnamefont{M.}~\bibnamefont{Wenzel}},
  \bibinfo{author}{\bibfnamefont{C.}~\bibnamefont{Wink}},
  \bibinfo{author}{\bibfnamefont{T.}~\bibnamefont{Maier}},
  \bibinfo{author}{\bibfnamefont{I.}~\bibnamefont{Ferrier-Barbut}},
  \bibnamefont{and} \bibinfo{author}{\bibfnamefont{T.}~\bibnamefont{Pfau}},
  \bibinfo{journal}{Nature} \textbf{\bibinfo{volume}{530}},
  \bibinfo{pages}{194} (\bibinfo{year}{2016}).

\bibitem[{\citenamefont{Ferrier-Barbut
  et~al.}(2016)\citenamefont{Ferrier-Barbut, Kadau, Schmitt, Wenzel, and
  Pfau}}]{expt_dy16_1}
\bibinfo{author}{\bibfnamefont{I.}~\bibnamefont{Ferrier-Barbut}},
  \bibinfo{author}{\bibfnamefont{H.}~\bibnamefont{Kadau}},
  \bibinfo{author}{\bibfnamefont{M.}~\bibnamefont{Schmitt}},
  \bibinfo{author}{\bibfnamefont{M.}~\bibnamefont{Wenzel}}, \bibnamefont{and}
  \bibinfo{author}{\bibfnamefont{T.}~\bibnamefont{Pfau}},
  \bibinfo{journal}{Phys. Rev. Lett.} \textbf{\bibinfo{volume}{116}},
  \bibinfo{pages}{215301} (\bibinfo{year}{2016}).

\bibitem[{\citenamefont{Chomaz et~al.}(2016)\citenamefont{Chomaz, Baier,
  Petter, Mark, W\"achtler, Santos, and Ferlaino}}]{expt_er16}
\bibinfo{author}{\bibfnamefont{L.}~\bibnamefont{Chomaz}},
  \bibinfo{author}{\bibfnamefont{S.}~\bibnamefont{Baier}},
  \bibinfo{author}{\bibfnamefont{D.}~\bibnamefont{Petter}},
  \bibinfo{author}{\bibfnamefont{J.}~\bibnamefont{Mark}, \bibfnamefont{M}},
  \bibinfo{author}{\bibfnamefont{F.}~\bibnamefont{W\"achtler}},
  \bibinfo{author}{\bibfnamefont{L.}~\bibnamefont{Santos}}, \bibnamefont{and}
  \bibinfo{author}{\bibfnamefont{F.}~\bibnamefont{Ferlaino}},
  \bibinfo{journal}{arXiv:1607.06613}  (\bibinfo{year}{2016}).

\bibitem[{\citenamefont{Schmitt et~al.}(2016)\citenamefont{Schmitt, Wenzel,
  B\"ottcher, Ferrier-Barbut, and Pfau}}]{expt_dy16_2}
\bibinfo{author}{\bibfnamefont{M.}~\bibnamefont{Schmitt}},
  \bibinfo{author}{\bibfnamefont{M.}~\bibnamefont{Wenzel}},
  \bibinfo{author}{\bibfnamefont{F.}~\bibnamefont{B\"ottcher}},
  \bibinfo{author}{\bibfnamefont{I.}~\bibnamefont{Ferrier-Barbut}},
  \bibnamefont{and} \bibinfo{author}{\bibfnamefont{T.}~\bibnamefont{Pfau}},
  \bibinfo{journal}{arXiv:1607.07355}  (\bibinfo{year}{2016}).

\bibitem[{\citenamefont{Lima and Pelster}(2011)}]{lima_Qfl_11}
\bibinfo{author}{\bibfnamefont{A.~R.~P.} \bibnamefont{Lima}} \bibnamefont{and}
  \bibinfo{author}{\bibfnamefont{A.}~\bibnamefont{Pelster}},
  \bibinfo{journal}{Phys. Rev. A} \textbf{\bibinfo{volume}{84}},
  \bibinfo{pages}{041604} (\bibinfo{year}{2011}).

\bibitem[{\citenamefont{W\"achtler and Santos}(2016{\natexlab{a}})}]{wac_fil16}
\bibinfo{author}{\bibfnamefont{F.}~\bibnamefont{W\"achtler}} \bibnamefont{and}
  \bibinfo{author}{\bibfnamefont{L.}~\bibnamefont{Santos}},
  \bibinfo{journal}{Phys. Rev. A} \textbf{\bibinfo{volume}{93}},
  \bibinfo{pages}{061603} (\bibinfo{year}{2016}{\natexlab{a}}).

\bibitem[{\citenamefont{Bisset et~al.}(2016)\citenamefont{Bisset, Wilson,
  Baillie, and Blakie}}]{bis_fl_16}
\bibinfo{author}{\bibfnamefont{N.}~\bibnamefont{Bisset}, \bibfnamefont{R}},
  \bibinfo{author}{\bibfnamefont{M.}~\bibnamefont{Wilson}, \bibfnamefont{R}},
  \bibinfo{author}{\bibfnamefont{D.}~\bibnamefont{Baillie}}, \bibnamefont{and}
  \bibinfo{author}{\bibfnamefont{B.}~\bibnamefont{Blakie}, \bibfnamefont{P}},
  \bibinfo{journal}{arXiv:1605.04964}  (\bibinfo{year}{2016}).

\bibitem[{\citenamefont{W\"achtler and
  Santos}(2016{\natexlab{b}})}]{wach_pro_16}
\bibinfo{author}{\bibfnamefont{F.}~\bibnamefont{W\"achtler}} \bibnamefont{and}
  \bibinfo{author}{\bibfnamefont{L.}~\bibnamefont{Santos}},
  \bibinfo{journal}{arXiv:1605.08676}  (\bibinfo{year}{2016}{\natexlab{b}}).

\bibitem[{\citenamefont{Baillie et~al.}(2016)\citenamefont{Baillie, Wilson,
  Bisset, and Blakie}}]{bal_qf_16}
\bibinfo{author}{\bibfnamefont{D.}~\bibnamefont{Baillie}},
  \bibinfo{author}{\bibfnamefont{M.}~\bibnamefont{Wilson}, \bibfnamefont{R}},
  \bibinfo{author}{\bibfnamefont{N.}~\bibnamefont{Bisset}, \bibfnamefont{R}},
  \bibnamefont{and} \bibinfo{author}{\bibfnamefont{B.}~\bibnamefont{Blakie},
  \bibfnamefont{P}}, \bibinfo{journal}{arXiv:1606.00824}
  (\bibinfo{year}{2016}).

\bibitem[{\citenamefont{Santos et~al.}(2003)\citenamefont{Santos, Shlyapnikov,
  and Lewenstein}}]{san_rot03}
\bibinfo{author}{\bibfnamefont{L.}~\bibnamefont{Santos}},
  \bibinfo{author}{\bibfnamefont{G.~V.} \bibnamefont{Shlyapnikov}},
  \bibnamefont{and}
  \bibinfo{author}{\bibfnamefont{M.}~\bibnamefont{Lewenstein}},
  \bibinfo{journal}{Phys. Rev. Lett.} \textbf{\bibinfo{volume}{90}},
  \bibinfo{pages}{250403} (\bibinfo{year}{2003}).

\bibitem[{\citenamefont{Pedri and Santos}(2005)}]{sol-luis}
\bibinfo{author}{\bibfnamefont{P.}~\bibnamefont{Pedri}} \bibnamefont{and}
  \bibinfo{author}{\bibfnamefont{L.}~\bibnamefont{Santos}},
  \bibinfo{journal}{Phys. Rev. Lett.} \textbf{\bibinfo{volume}{95}},
  \bibinfo{pages}{200404} (\bibinfo{year}{2005}).

\bibitem[{\citenamefont{Tikhonenkov et~al.}(2008)\citenamefont{Tikhonenkov,
  Malomed, and Vardi}}]{sol-aniso}
\bibinfo{author}{\bibfnamefont{I.}~\bibnamefont{Tikhonenkov}},
  \bibinfo{author}{\bibfnamefont{B.~A.} \bibnamefont{Malomed}},
  \bibnamefont{and} \bibinfo{author}{\bibfnamefont{A.}~\bibnamefont{Vardi}},
  \bibinfo{journal}{Phys. Rev. Lett.} \textbf{\bibinfo{volume}{100}},
  \bibinfo{pages}{090406} (\bibinfo{year}{2008}).

\bibitem[{\citenamefont{Ronen et~al.}(2007)\citenamefont{Ronen, Bortolotti, and
  Bohn}}]{shai_rot_07}
\bibinfo{author}{\bibfnamefont{S.}~\bibnamefont{Ronen}},
  \bibinfo{author}{\bibfnamefont{D.~C.~E.} \bibnamefont{Bortolotti}},
  \bibnamefont{and} \bibinfo{author}{\bibfnamefont{J.~L.} \bibnamefont{Bohn}},
  \bibinfo{journal}{Phys. Rev. Lett.} \textbf{\bibinfo{volume}{98}},
  \bibinfo{pages}{030406} (\bibinfo{year}{2007}).

\bibitem[{\citenamefont{Fedorov et~al.}(2014)\citenamefont{Fedorov, Kurbakov,
  Shchadilova, and Lozovik}}]{td_fed_14}
\bibinfo{author}{\bibfnamefont{A.~K.} \bibnamefont{Fedorov}},
  \bibinfo{author}{\bibfnamefont{I.~L.} \bibnamefont{Kurbakov}},
  \bibinfo{author}{\bibfnamefont{Y.~E.} \bibnamefont{Shchadilova}},
  \bibnamefont{and} \bibinfo{author}{\bibfnamefont{Y.~E.}
  \bibnamefont{Lozovik}}, \bibinfo{journal}{Phys. Rev. A}
  \textbf{\bibinfo{volume}{90}}, \bibinfo{pages}{043616}
  (\bibinfo{year}{2014}).

\bibitem[{\citenamefont{Baillie and Blakie}(2015)}]{bai_til_15}
\bibinfo{author}{\bibfnamefont{D.}~\bibnamefont{Baillie}} \bibnamefont{and}
  \bibinfo{author}{\bibfnamefont{P.~B.} \bibnamefont{Blakie}},
  \bibinfo{journal}{New J. Phys.} \textbf{\bibinfo{volume}{17}},
  \bibinfo{pages}{033028} (\bibinfo{year}{2015}).

\bibitem[{\citenamefont{Mulkerin et~al.}(2013)\citenamefont{Mulkerin, van
  Bijnen, O'Dell, Martin, and Parker}}]{mul_til_vor13}
\bibinfo{author}{\bibfnamefont{B.~C.} \bibnamefont{Mulkerin}},
  \bibinfo{author}{\bibfnamefont{R.~M.~W.} \bibnamefont{van Bijnen}},
  \bibinfo{author}{\bibfnamefont{D.~H.~J.} \bibnamefont{O'Dell}},
  \bibinfo{author}{\bibfnamefont{A.~M.} \bibnamefont{Martin}},
  \bibnamefont{and} \bibinfo{author}{\bibfnamefont{N.~G.}
  \bibnamefont{Parker}}, \bibinfo{journal}{Phys. Rev. Lett.}
  \textbf{\bibinfo{volume}{111}}, \bibinfo{pages}{170402}
  (\bibinfo{year}{2013}).

\bibitem[{\citenamefont{Edmonds et~al.}(2016)\citenamefont{Edmonds, Bland,
  O'Dell, and Parker}}]{ds_tilt_16}
\bibinfo{author}{\bibfnamefont{M.~J.} \bibnamefont{Edmonds}},
  \bibinfo{author}{\bibfnamefont{T.}~\bibnamefont{Bland}},
  \bibinfo{author}{\bibfnamefont{D.~H.~J.} \bibnamefont{O'Dell}},
  \bibnamefont{and} \bibinfo{author}{\bibfnamefont{N.~G.}
  \bibnamefont{Parker}}, \bibinfo{journal}{Phys. Rev. A}
  \textbf{\bibinfo{volume}{93}}, \bibinfo{pages}{063617}
  (\bibinfo{year}{2016}).

\bibitem[{\citenamefont{Bismut et~al.}(2012)\citenamefont{Bismut,
  Laburthe-Tolra, Mar\'echal, Pedri, Gorceix, and Vernac}}]{bis_ani_sp12}
\bibinfo{author}{\bibfnamefont{G.}~\bibnamefont{Bismut}},
  \bibinfo{author}{\bibfnamefont{B.}~\bibnamefont{Laburthe-Tolra}},
  \bibinfo{author}{\bibfnamefont{E.}~\bibnamefont{Mar\'echal}},
  \bibinfo{author}{\bibfnamefont{P.}~\bibnamefont{Pedri}},
  \bibinfo{author}{\bibfnamefont{O.}~\bibnamefont{Gorceix}}, \bibnamefont{and}
  \bibinfo{author}{\bibfnamefont{L.}~\bibnamefont{Vernac}},
  \bibinfo{journal}{Phys. Rev. Lett.} \textbf{\bibinfo{volume}{109}},
  \bibinfo{pages}{155302} (\bibinfo{year}{2012}).

\bibitem[{\citenamefont{Ticknor et~al.}(2011)\citenamefont{Ticknor, Wilson, and
  Bohn}}]{tic_Asf_11}
\bibinfo{author}{\bibfnamefont{C.}~\bibnamefont{Ticknor}},
  \bibinfo{author}{\bibfnamefont{R.~M.} \bibnamefont{Wilson}},
  \bibnamefont{and} \bibinfo{author}{\bibfnamefont{J.~L.} \bibnamefont{Bohn}},
  \bibinfo{journal}{Phys. Rev. Lett.} \textbf{\bibinfo{volume}{106}},
  \bibinfo{pages}{065301} (\bibinfo{year}{2011}).

\bibitem[{\citenamefont{Zhang et~al.}(2015)\citenamefont{Zhang, Safavi-Naini,
  Rey, and Capogrosso-Sansone}}]{zha_til_10}
\bibinfo{author}{\bibfnamefont{C.}~\bibnamefont{Zhang}},
  \bibinfo{author}{\bibfnamefont{A.}~\bibnamefont{Safavi-Naini}},
  \bibinfo{author}{\bibfnamefont{A.~M.} \bibnamefont{Rey}}, \bibnamefont{and}
  \bibinfo{author}{\bibfnamefont{B.}~\bibnamefont{Capogrosso-Sansone}},
  \bibinfo{journal}{New J. Phys.} \textbf{\bibinfo{volume}{17}},
  \bibinfo{pages}{123014} (\bibinfo{year}{2015}).

\bibitem[{\citenamefont{Macia et~al.}(2014)\citenamefont{Macia, Boronat, and
  Mazzanti}}]{mac_til_14}
\bibinfo{author}{\bibfnamefont{A.}~\bibnamefont{Macia}},
  \bibinfo{author}{\bibfnamefont{J.}~\bibnamefont{Boronat}}, \bibnamefont{and}
  \bibinfo{author}{\bibfnamefont{F.}~\bibnamefont{Mazzanti}},
  \bibinfo{journal}{Phys. Rev. A} \textbf{\bibinfo{volume}{90}},
  \bibinfo{pages}{061601} (\bibinfo{year}{2014}).

\bibitem[{\citenamefont{Wu et~al.}(2016)\citenamefont{Wu, Block, and
  Bruun}}]{wu_lcry_16}
\bibinfo{author}{\bibfnamefont{Z.}~\bibnamefont{Wu}},
  \bibinfo{author}{\bibfnamefont{J.~K.} \bibnamefont{Block}}, \bibnamefont{and}
  \bibinfo{author}{\bibfnamefont{G.~M.} \bibnamefont{Bruun}},
  \bibinfo{journal}{Scientific Reports} \textbf{\bibinfo{volume}{6}},
  \bibinfo{pages}{19038 EP } (\bibinfo{year}{2016}).

\bibitem[{\citenamefont{Macia et~al.}(2012)\citenamefont{Macia, Hufnagl,
  Mazzanti, Boronat, and Zillich}}]{mac_str_12}
\bibinfo{author}{\bibfnamefont{A.}~\bibnamefont{Macia}},
  \bibinfo{author}{\bibfnamefont{D.}~\bibnamefont{Hufnagl}},
  \bibinfo{author}{\bibfnamefont{F.}~\bibnamefont{Mazzanti}},
  \bibinfo{author}{\bibfnamefont{J.}~\bibnamefont{Boronat}}, \bibnamefont{and}
  \bibinfo{author}{\bibfnamefont{R.~E.} \bibnamefont{Zillich}},
  \bibinfo{journal}{Phys. Rev. Lett.} \textbf{\bibinfo{volume}{109}},
  \bibinfo{pages}{235307} (\bibinfo{year}{2012}).

\bibitem[{\citenamefont{Khaykovich et~al.}(2002)\citenamefont{Khaykovich,
  Schreck, Ferrari, Bourdel, Cubizolles, Carr, Castin, and
  Salomon}}]{sol_expt_02}
\bibinfo{author}{\bibfnamefont{L.}~\bibnamefont{Khaykovich}},
  \bibinfo{author}{\bibfnamefont{F.}~\bibnamefont{Schreck}},
  \bibinfo{author}{\bibfnamefont{G.}~\bibnamefont{Ferrari}},
  \bibinfo{author}{\bibfnamefont{T.}~\bibnamefont{Bourdel}},
  \bibinfo{author}{\bibfnamefont{J.}~\bibnamefont{Cubizolles}},
  \bibinfo{author}{\bibfnamefont{L.~D.} \bibnamefont{Carr}},
  \bibinfo{author}{\bibfnamefont{Y.}~\bibnamefont{Castin}}, \bibnamefont{and}
  \bibinfo{author}{\bibfnamefont{C.}~\bibnamefont{Salomon}},
  \bibinfo{journal}{Science} \textbf{\bibinfo{volume}{296}},
  \bibinfo{pages}{1290} (\bibinfo{year}{2002}).

\bibitem[{\citenamefont{Strecker et~al.}(2002)\citenamefont{Strecker,
  Partridge, Truscott, and Hulet}}]{sol_expt_Nat_02}
\bibinfo{author}{\bibfnamefont{K.~E.} \bibnamefont{Strecker}},
  \bibinfo{author}{\bibfnamefont{G.~B.} \bibnamefont{Partridge}},
  \bibinfo{author}{\bibfnamefont{A.~G.} \bibnamefont{Truscott}},
  \bibnamefont{and} \bibinfo{author}{\bibfnamefont{R.~G.} \bibnamefont{Hulet}},
  \bibinfo{journal}{Nature} \textbf{\bibinfo{volume}{417}},
  \bibinfo{pages}{150} (\bibinfo{year}{2002}).

\bibitem[{\citenamefont{Cornish et~al.}(2006)\citenamefont{Cornish, Thompson,
  and Wieman}}]{sol_expt_06}
\bibinfo{author}{\bibfnamefont{S.~L.} \bibnamefont{Cornish}},
  \bibinfo{author}{\bibfnamefont{S.~T.} \bibnamefont{Thompson}},
  \bibnamefont{and} \bibinfo{author}{\bibfnamefont{C.~E.}
  \bibnamefont{Wieman}}, \bibinfo{journal}{Phys. Rev. Lett.}
  \textbf{\bibinfo{volume}{96}}, \bibinfo{pages}{170401}
  (\bibinfo{year}{2006}).

\bibitem[{\citenamefont{Parker and O'Dell}(2008)}]{par_TF08}
\bibinfo{author}{\bibfnamefont{N.~G.} \bibnamefont{Parker}} \bibnamefont{and}
  \bibinfo{author}{\bibfnamefont{D.~H.~J.} \bibnamefont{O'Dell}},
  \bibinfo{journal}{Phys. Rev. A} \textbf{\bibinfo{volume}{78}},
  \bibinfo{pages}{041601} (\bibinfo{year}{2008}).

\bibitem[{\citenamefont{Nath et~al.}(2009)\citenamefont{Nath, Pedri, and
  Santos}}]{nath-pho}
\bibinfo{author}{\bibfnamefont{R.}~\bibnamefont{Nath}},
  \bibinfo{author}{\bibfnamefont{P.}~\bibnamefont{Pedri}}, \bibnamefont{and}
  \bibinfo{author}{\bibfnamefont{L.}~\bibnamefont{Santos}},
  \bibinfo{journal}{Phys. Rev. Lett.} \textbf{\bibinfo{volume}{102}},
  \bibinfo{pages}{050401} (\bibinfo{year}{2009}).

\bibitem[{\citenamefont{Mulkerin et~al.}(2014)\citenamefont{Mulkerin, O'Dell,
  Martin, and Parker}}]{mul_vor_til14}
\bibinfo{author}{\bibfnamefont{B.~C.} \bibnamefont{Mulkerin}},
  \bibinfo{author}{\bibfnamefont{D.~H.~J.} \bibnamefont{O'Dell}},
  \bibinfo{author}{\bibfnamefont{A.~M.} \bibnamefont{Martin}},
  \bibnamefont{and} \bibinfo{author}{\bibfnamefont{N.~G.}
  \bibnamefont{Parker}}, \bibinfo{journal}{Journal of Physics: Conference
  Series} \textbf{\bibinfo{volume}{497}}, \bibinfo{pages}{012025}
  (\bibinfo{year}{2014}).

\bibitem[{\citenamefont{Bruun and Taylor}(2008)}]{bru_fer_08}
\bibinfo{author}{\bibfnamefont{G.~M.} \bibnamefont{Bruun}} \bibnamefont{and}
  \bibinfo{author}{\bibfnamefont{E.}~\bibnamefont{Taylor}},
  \bibinfo{journal}{Phys. Rev. Lett.} \textbf{\bibinfo{volume}{101}},
  \bibinfo{pages}{245301} (\bibinfo{year}{2008}).

\bibitem[{\citenamefont{Raghunandan et~al.}(2015)\citenamefont{Raghunandan,
  Mishra, \L{}akomy, Pedri, Santos, and Nath}}]{meg15}
\bibinfo{author}{\bibfnamefont{M.}~\bibnamefont{Raghunandan}},
  \bibinfo{author}{\bibfnamefont{C.}~\bibnamefont{Mishra}},
  \bibinfo{author}{\bibfnamefont{K.}~\bibnamefont{\L{}akomy}},
  \bibinfo{author}{\bibfnamefont{P.}~\bibnamefont{Pedri}},
  \bibinfo{author}{\bibfnamefont{L.}~\bibnamefont{Santos}}, \bibnamefont{and}
  \bibinfo{author}{\bibfnamefont{R.}~\bibnamefont{Nath}},
  \bibinfo{journal}{Phys. Rev. A} \textbf{\bibinfo{volume}{92}},
  \bibinfo{pages}{013637} (\bibinfo{year}{2015}).

\bibitem[{\citenamefont{Nath and Santos}(2010)}]{nath_far_10}
\bibinfo{author}{\bibfnamefont{R.}~\bibnamefont{Nath}} \bibnamefont{and}
  \bibinfo{author}{\bibfnamefont{L.}~\bibnamefont{Santos}},
  \bibinfo{journal}{Phys. Rev. A} \textbf{\bibinfo{volume}{81}},
  \bibinfo{pages}{033626} (\bibinfo{year}{2010}).

\bibitem[{\citenamefont{Fischer}(2006)}]{uwe_dip_06}
\bibinfo{author}{\bibfnamefont{U.~R.} \bibnamefont{Fischer}},
  \bibinfo{journal}{Phys. Rev. A} \textbf{\bibinfo{volume}{73}},
  \bibinfo{pages}{031602} (\bibinfo{year}{2006}).

\bibitem[{\citenamefont{Koch et~al.}(2008)\citenamefont{Koch, Lahaye, Metz,
  Frohlich, Griesmaier, and Pfau}}]{koc_nat_08}
\bibinfo{author}{\bibfnamefont{T.}~\bibnamefont{Koch}},
  \bibinfo{author}{\bibfnamefont{T.}~\bibnamefont{Lahaye}},
  \bibinfo{author}{\bibfnamefont{J.}~\bibnamefont{Metz}},
  \bibinfo{author}{\bibfnamefont{B.}~\bibnamefont{Frohlich}},
  \bibinfo{author}{\bibfnamefont{A.}~\bibnamefont{Griesmaier}},
  \bibnamefont{and} \bibinfo{author}{\bibfnamefont{T.}~\bibnamefont{Pfau}},
  \bibinfo{journal}{Nat Phys} \textbf{\bibinfo{volume}{4}},
  \bibinfo{pages}{218} (\bibinfo{year}{2008}).

\bibitem[{\citenamefont{P\'erez-Garc\'ia
  et~al.}(1996)\citenamefont{P\'erez-Garc\'ia, Michinel, Cirac, Lewenstein, and
  Zoller}}]{var_zoll_96}
\bibinfo{author}{\bibfnamefont{V.~M.} \bibnamefont{P\'erez-Garc\'ia}},
  \bibinfo{author}{\bibfnamefont{H.}~\bibnamefont{Michinel}},
  \bibinfo{author}{\bibfnamefont{J.~I.} \bibnamefont{Cirac}},
  \bibinfo{author}{\bibfnamefont{M.}~\bibnamefont{Lewenstein}},
  \bibnamefont{and} \bibinfo{author}{\bibfnamefont{P.}~\bibnamefont{Zoller}},
  \bibinfo{journal}{Phys. Rev. Lett.} \textbf{\bibinfo{volume}{77}},
  \bibinfo{pages}{5320} (\bibinfo{year}{1996}).

\bibitem[{\citenamefont{Yi and You}(2001)}]{var_you_01}
\bibinfo{author}{\bibfnamefont{S.}~\bibnamefont{Yi}} \bibnamefont{and}
  \bibinfo{author}{\bibfnamefont{L.}~\bibnamefont{You}},
  \bibinfo{journal}{Phys. Rev. A} \textbf{\bibinfo{volume}{63}},
  \bibinfo{pages}{053607} (\bibinfo{year}{2001}).

\end{thebibliography}
\end{document}